\documentclass[article]{aastex631}

\shorttitle{Many Headed Hydra Modal Basis}
\shortauthors{Fowler \& Stelter}
\graphicspath{{./}{figures/}}

\begin{document}

\title{Wavefront Sensing and Control with the Many Headed Hydra Modal Basis}

\author[0000-0002-0726-9323]{J. Fowler}
\affiliation{University of California, Santa Cruz \\
1156 High St \\
Santa Cruz, CA 95064, USA}

\author[0000-0003-4549-0210-]{R. Deno Stelter}
\affiliation{UC Observatories \\
University of California, Santa Cruz \\
1156 High St \\
Santa Cruz, CA, 95060, USA}



\begin{abstract}

The future of space and ground based telescopes is intimately tied to technology and algorithm development surrounding wavefront sensing and control. Only with cutting edge developments and unusual ideas will we be able to build diffraction limited observatories on the ground that contend with earth-atmosphere, as well as space-based observatories that sense and control for optical aberrations and telescope jitter. There exist a variety of mathematical bases for decomposing wavefront images, including the zonal modal basis, Zernike polynomials, and Fourier modes. However, previous bases have neglected the most vital element of astronomical optics: our field is only as good as the people in it. In this paper we propose a new Many Headed Hydra Modal Basis (hydra heads) for wavefront decomposition, using physical representations of the Adaptive Optics community. We find that the first author makes the best wavefront decomposition, on the order of a correct reconstruction within $\sim 1\%$ of the original turbulence image. We discuss engineering implications of the Many Headed Hydra Modal Basis, as applied to deformable mirror technology and within active and adaptive optics control loops. Finally, we explore predictive control avenues, by positing that the first author being the most effective wavefront predicts a successful future for them as a high contrast imaging scientist. 

\end{abstract}


\section{Introduction} \label{sec:intro}
Future advances in astrophysics depend on improving performance our our telescopes and instruments, ranging from verifying sub-Earth exoplanets around surprisingly nearby stars~\citep{IndependentDiscoveryOfSubEarth}, to identifying floofs at otpical wavelengths~\citep{RotationalFloofs}, to discovering exigent threats in galaxies far, far away~\citep{UltimatePower}. The performance of these instruments is inextricably entwined with how well we are able to sense and correct for aberrations in our incoming light -- be it from earth-atmosphere, errors built into the telescope system, or often both. Given we start with some representation of phase from an incoming wavefront, and hope to correct it, the next step in this sensing and control problem is a choice of basis. A zonal modal basis \citep{zonal} is intuitively simple -- breaking down each wavefront sensor subaperture into its own mode and controlling along zones. Whereas, a Zernike modal basis \citep{zernike} builds up an intuition of optical aberrations, framing incoming light in terms of coma, astigmatism, trefoil, etc, that maps to optical defects within our system. Finally, a Fourier modal basis \citep{fourier} decomposes light into sine and cosine waves, and has a natural connection to Fourier transforms into frequency space. 

We present the Many Headed Hydra Modal Basis (hydra heads), as a basis that encapsulates the true character of adaptive optics, by encoding the likeness of the very scientists who do it a basis sets. With Section \ref{sec:hydra-head}, we work through the formalism of the hydra heads. In Section \ref{sec:efficacy} we examine the effectiveness of the hydra heads as used to decompose a sample turbulence image. In Section \ref{sec:control}, we discuss wavefront control applications of the hydra heads. Finally, in Section \ref{sec:conclusion} we make our final conclusions about why the first author makes an extremely good basis set.

\section{Formulation of the Many Headed Hydra Modal Basis} \label{sec:hydra-head}

Given 16 starting candidate heads, we build the modes by:
\begin{enumerate}
    \item Making the head into a square image. 
    \item Converting the image from color to black and white. 
    \item Rescaling the image from 0-1. 
    \item Appending a negative valued copy of the image. 
    \item Increasing the spatial frequency (number of heads) and altering the orientation. 
\end{enumerate}

\begin{figure*}
\gridline{\fig{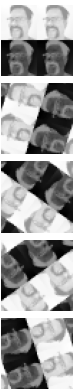}{0.2\textwidth}{}
          \fig{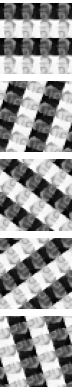}{0.2\textwidth}{}
          \fig{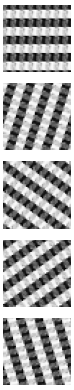}{0.2\textwidth}{}
          \fig{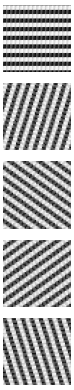}{0.2\textwidth}{}
          \fig{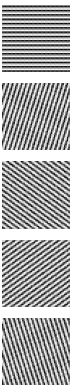}{0.2\textwidth}{}
          }
\caption{Example Many Headed Hydra Modal Basis set using a Debes head as starting input. While this full set has 2500 modes (spanning tighter degree increments as the pattern rotates and more repetitions of heads) we display a handful of modes to get an idea of the modal basis images.
\label{fig:debes}}
\end{figure*}

We display one sample hydra head in Figure \ref{fig:debes}, built with the Debes head. 

Appendix \ref{sec:hydrae-sim} demonstrates an example the \href{https://github.com/julesfowler/hydra-sim}{\texttt{hydra-sim}} software used to create the hydra heads. Appendix \ref{sec:alltheheads} scans through a few examples of modes for each input head.

\section{Efficacy Among the Hydra Heads} \label{sec:efficacy}

Given an input set of 16 heads, we built a Many Headed Hydra Modal Basis for each head, as described in Section \ref{sec:hydra-head}. With this suite of hydrae, we can then test them practically, by decomposing a sample turbulence image. 
Using \texttt{HCIPy} \citep{hcipy}, we simulate a turbulence profile with 5 wind layers, each with their own velocity and $r_0$. Table \ref{tab:atmosim} outlines the atmospheric parameters used to simulate this atmosphere. 

\begin{deluxetable*}{llll}
\tablenum{1}
\tablecaption{Atmospheric turbulence simulation. \label{tab:atmosim}}
\tablewidth{0pt}
\tablehead{
\colhead{Layer} & \colhead{Velocity} & \colhead{Direction} & \colhead{$r_0$} \\
 & \colhead{m/s} & \colhead{degrees} & \colhead{cm} }
\startdata
1 & 22.7 & 246 & 38.9 \\
2 & 3.28 & 71 & 44.7 \\
3 & 16.6 & 294 & 45.5 \\
4 & 5.89 & 150 & 38.8 \\
5 & 19.8 & 14 & 43.6 \\
\enddata
\tablecomments{These conditions were used to simulate an atmospheric turbulence profile as a practical test of the hydrae. We used conditions from \citet{poyneer2007}, because the first author happened to have simulated this atmosphere at high resolution recently. Our final images are simulated at a wavelength of 1630 nm over an 8 meter primary telescope mirror.}
\end{deluxetable*}

Given each hydra head mode starts as a square m by m matrix, which could map to a representation of a turbulence profile binned down to a resolution of m x m, we make the Many Headed Hydra Modal Basis set $\mathbf{A}$, by flattening each mode to be a column of $\mathbf{A}$. From there, decomposing a turbulence image becomes a linear algebra problem of the form: 
\begin{equation}
\mathbf{A}\vec{x} = \vec{b}
\end{equation}

where $\vec{b}$ is the flattened image we aim to reconstruct, and $\vec{x}$ contains the coefficients of each mode to reconstruct the given image. We solve for $\vec{x}$ using a pseudo-inversion of $\mathbf{A}$, using \texttt{numpy}'s singular value decomposition pseduo-inverse method. 

\begin{figure}
\plotone{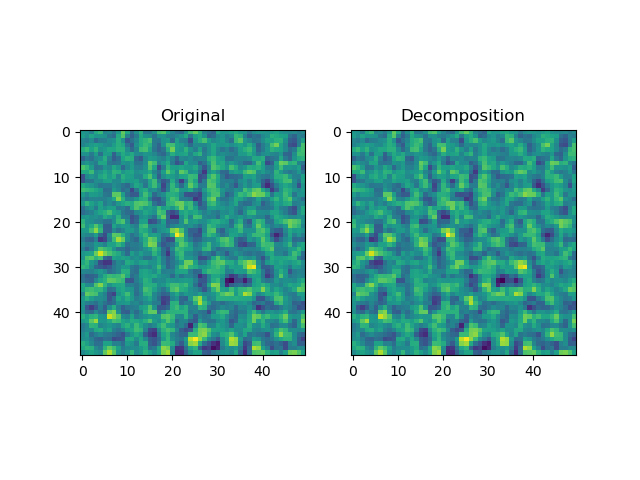}
\caption{Left: Original scene of turbulence, originally generated at a resolution of 100 pixels under the conditions described in \ref{tab:atmosim}, and binned down to 50. Right: Reconstruction using the Fowler hydra head. 
\label{fig:turb_images}}
\end{figure}

\begin{figure}
\plotone{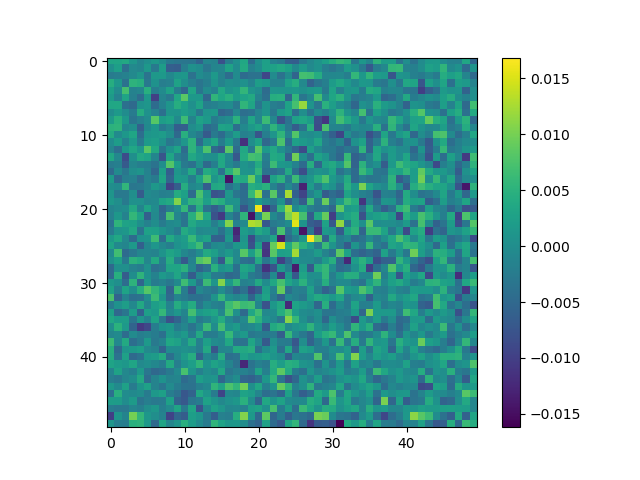}
\caption{Difference image between the Fowler hydra head reconstruction and the original turbulence. Not the scale on the order of 0.01 radians in the infrared regime.
\label{fig:difference}}
\end{figure}

For this test we generated 2500 modes for each head, each a 50x50 scene, which was then used to decompose an image simulated at 100x100 resolution and binned down to match the 50x50 (to preserve Nyquist sampling). Table \ref{tab:heads} shows the relative performance of each hydra head. While the Hirschauer head displays the best performance, a closer inspection of the original input images reveals that the Hirschauer image benefits from a flat background. Considering the unreliability of dome flats, we rule this out as a potential hydra head, and find that the Fowler hydra head is the most effective modal basis in this set. Figure \ref{fig:turb_images} shows an example turbulence image and the reconstruction using the Fowler hydra head. Figure \ref{fig:difference} shows the difference image between the reconstruction and the original turbulence. 

\begin{deluxetable*}{lc}
\tablenum{1}
\tablecaption{Comparative performance of the hydra heads. \label{tab:heads}}
\tablewidth{0pt}
\tablehead{
\colhead{Head} & \colhead{$\%$ Error}
}
\startdata
\textit{Hirschauer}& 0.09 \\ 
Fowler& 1.07 \\ 
Hinz  & 1.54 \\
Soummer & 2.47 \\
van Belle  & 2.81 \\
Kirkpatrick  &  3.81 \\
Macintosh  & 3.89 \\
Pearce  & 4.93 \\
Stelter & 5.02 \\
Jensen-Clem  & 5.33 \\
Ngo  & 5.57 \\
Debes  & 5.75 \\
Bowens-Rubin  & 11.97 \\
Steiger  & 12.34 \\
Lewis  & 13.92 \\
Balmer  & 15.66 \\
\enddata
\tablecomments{The relative performance of the sample heads. We decompose a sample turbulent wavefront with a hydra head, and calculate the percent difference in the reconstructed image and the original simulated image. }
\end{deluxetable*}

\section{Control Applications} \label{sec:control}

The hydrae are well-suited for pupils produced by off-axis mirrors (such as OAPs or OAEs) because the pupil planes of off-axis mirrors are tilted with respect to the beam direction resulting in an elongated pupil.
The pupil, being non-square, is thus well-matched to a craftily-cropped head shot, although care must be taken to ensure that the head shot and the pupil share the same axial length ratio.
Deviations can be systematically addressed by adding more fractal layers, which reduces the problem to one where the solution is known.

While head shots of prominent scientists\footnote{
Or at least the ones who responded to a late-night call on Twitter for head shots.}
may not, at first, seem to be a sensible basis for wavefront control, we note that, from the perspective of a deformable mirror (DM) actuator, following a leader is second nature.
After all, a `follower/leader' actuator scheme is a common choice in designing wavefront control loops.
We have simplified this scheme by finding the best head shot and designating it the leader, allowing the DM actuators to act in accordance with their natural predilections.

Edge effect artifacts of the head shots can affect the final result.
We have implemented a simple fractal-headed approach described below, simply allowing the higher frequency features to be fit with (spatially) copies of smaller head shots as needed.\footnote{
We expect that snapshots of stacks of turtles would also be effective.\citep{TurtlesJohnGreen}} 
Determining the lower end cut-off frequency scale when using the hydra heads requires the use of a carefully-timed application of an integrating sphere fed by a high-powered incandescent lamp or laser, or cryogenic liquid (we note that open flame light sources are considered more traditional).
With some trial and error, the suitable cut-off frequency can be found experimentally with only mild burns.

\section{Conclusions} \label{sec:conclusion}

In conclusion, we find that decomposing turbulence with pictures of astronomers is surprisingly feasible, with a shockingly high variability. While it seemed likely that this project would essentially test how good heads are at being sin waves, further inspection of the individual bases (see Appendix \ref{sec:alltheheads}) shows a striking similarity at higher order modes, as only a handful of pixels can be used to represent an image that requires many features to resemble a head. This potentially implies that the lower order modes are having a large impact on this decomposition. 

Finally, we stand by the conclusion that how well your head can decompose turbulence is strongly predictive of scientific output, with a focus on how well the first author can reconstruct the sample scene.\footnote{This effect has no relation to Fowler sampling (a commonly-used scheme for reading out infrared detectors), as far as we are aware.} No formal statistics have been conducted on this sample to verify this, and we do not recommend this for future works. 

\begin{acknowledgments}
Thanks to William Balmer, Rachel Bowens-Rubin, John Debes, Alec Hirschauer, Allison Kirkpatrick, Briley Lewis, Bruce Macintosh, Henry Ngo, Logan Pearce, Phil Hinz, Rebecca Jensen-Clem, Sarah Steiger, Gerard van Belle, and Remi Soummer for really putting their heads together to make this project possible. 
\end{acknowledgments}

\newpage

\software{\texttt{astropy} \citep{2018AJ....156..123A},  
          \texttt{matplotlib}, \citep{matplotlib},
          \texttt{numpy} \citep{2013RMxAA..49..137F}, 
          \texttt{PIL}, \citep{pil},
          \texttt{HCIPy}, \citep{hcipy},
          }



\appendix

\section{\texttt{hydraa-sim} Simulating the Performance of the Many Headed Hydra Modal Basis} \label{sec:hydrae-sim}

\begin{figure}[b]
\plotone{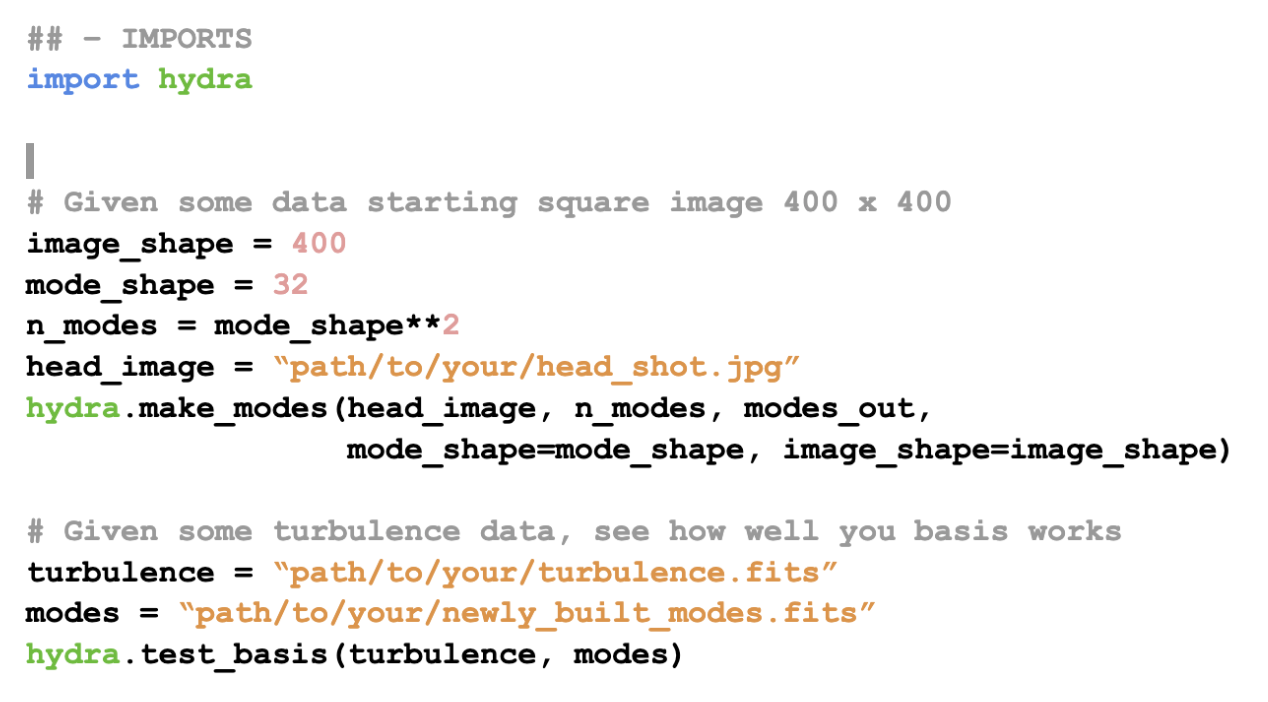}
\caption{Example code using \texttt{hydra-sim} to build a hydra head and decompose an image. 
\label{fig:hydra-sim}}
\end{figure}

With this paper, we've released \href{https://github.com/julesfowler/hydra-sim}{\texttt{hydra-sim}}, allowing any user to generate a modal basis from a square input image, and test it against a turbulence profile. Figure \ref{fig:hydra-sim} shows an example of using the code to build and test a modal basis. 

\newpage 

\section{Full Suite of Hydra Heads} \label{sec:alltheheads}

\subsection{Hirschauer Basis}
\begin{figure*}[b]
\gridline{\fig{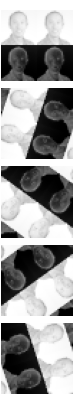}{0.2\textwidth}{}
          \fig{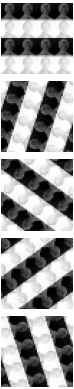}{0.2\textwidth}{}
          \fig{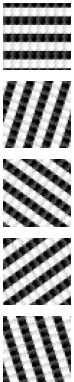}{0.2\textwidth}{}
          \fig{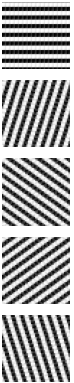}{0.2\textwidth}{}
          \fig{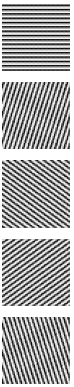}{0.2\textwidth}{}
          }
\caption{
\label{fig:hirschauer}}
\end{figure*}

\newpage

\subsection{Fowler Basis}
\begin{figure*}[b]
\gridline{\fig{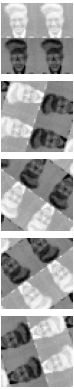}{0.2\textwidth}{}
          \fig{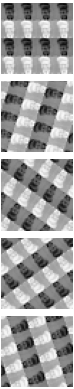}{0.2\textwidth}{}
          \fig{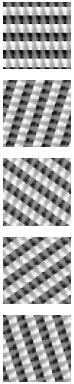}{0.2\textwidth}{}
          \fig{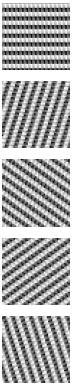}{0.2\textwidth}{}
          \fig{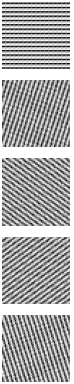}{0.2\textwidth}{}
          }
\caption{
\label{fig:fowler}}
\end{figure*}
\newpage

\subsection{Hinz Basis}
\begin{figure*}[b]
\gridline{\fig{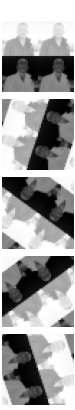}{0.2\textwidth}{}
          \fig{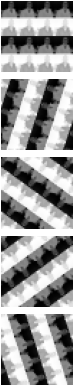}{0.2\textwidth}{}
          \fig{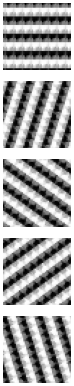}{0.2\textwidth}{}
          \fig{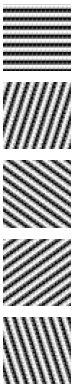}{0.2\textwidth}{}
          \fig{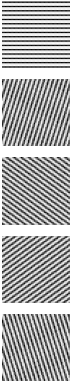}{0.2\textwidth}{}
          }
\caption{
\label{fig:hinz}}
\end{figure*}
\newpage

\subsection{Soummer Basis}
\begin{figure*}[b]
\gridline{\fig{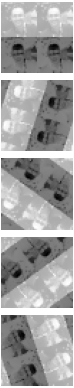}{0.2\textwidth}{}
          \fig{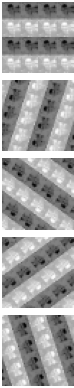}{0.2\textwidth}{}
          \fig{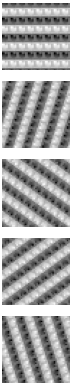}{0.2\textwidth}{}
          \fig{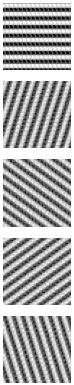}{0.2\textwidth}{}
          \fig{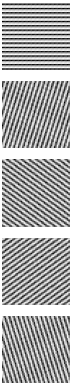}{0.2\textwidth}{}
          }
\caption{
\label{fig:soummer}}
\end{figure*}
\newpage

\subsection{van Belle Basis}[b]
\begin{figure*}[b]
\gridline{\fig{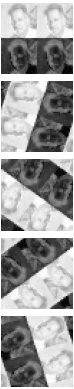}{0.2\textwidth}{}
          \fig{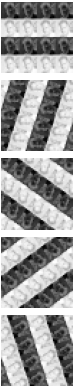}{0.2\textwidth}{}
          \fig{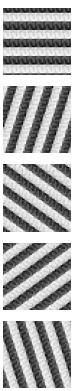}{0.2\textwidth}{}
          \fig{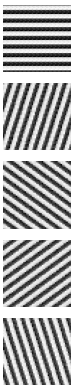}{0.2\textwidth}{}
          \fig{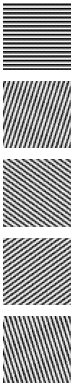}{0.2\textwidth}{}
          }
\caption{
\label{fig:vanbelle}}
\end{figure*}
\newpage

\subsection{Kirkpatrick Basis}
\begin{figure*}[b]
\gridline{\fig{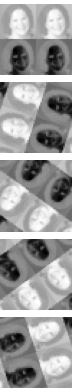}{0.2\textwidth}{}
          \fig{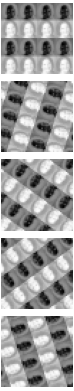}{0.2\textwidth}{}
          \fig{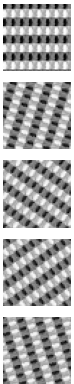}{0.2\textwidth}{}
          \fig{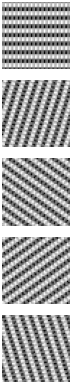}{0.2\textwidth}{}
          \fig{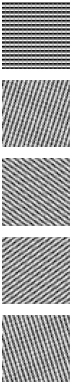}{0.2\textwidth}{}
          }
\caption{
\label{fig:kirkpatrick}}
\end{figure*}
\newpage

\subsection{Macintosh Basis}
\begin{figure*}[b]
\gridline{\fig{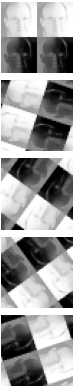}{0.2\textwidth}{}
          \fig{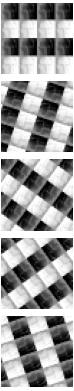}{0.2\textwidth}{}
          \fig{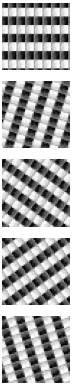}{0.2\textwidth}{}
          \fig{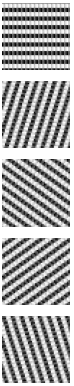}{0.2\textwidth}{}
          \fig{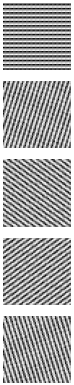}{0.2\textwidth}{}
          }
\caption{
\label{fig:macintosh}}
\end{figure*}
\newpage

\subsection{Pearce Basis}
\begin{figure*}[b]
\gridline{\fig{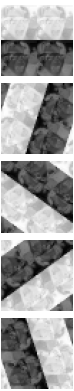}{0.2\textwidth}{}
          \fig{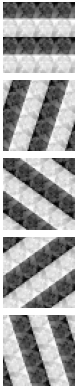}{0.2\textwidth}{}
          \fig{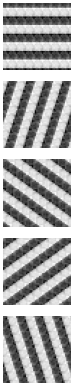}{0.2\textwidth}{}
          \fig{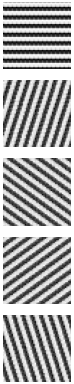}{0.2\textwidth}{}
          \fig{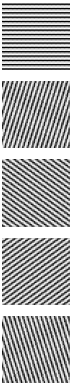}{0.2\textwidth}{}
          }
\caption{
\label{fig:pearce}}
\end{figure*}
\newpage

\subsection{Stelter Basis}
\begin{figure*}[b]
\gridline{\fig{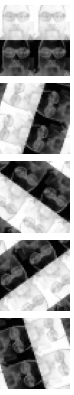}{0.2\textwidth}{}
          \fig{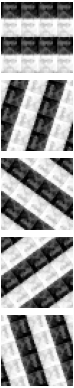}{0.2\textwidth}{}
          \fig{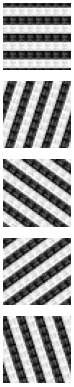}{0.2\textwidth}{}
          \fig{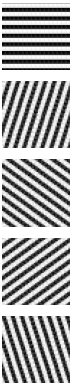}{0.2\textwidth}{}
          \fig{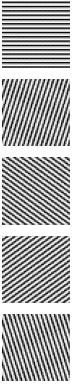}{0.2\textwidth}{}
          }
\caption{
\label{fig:stelter}}
\end{figure*}
\newpage

\subsection{Jensen-Clem Basis}
\begin{figure*}[b]
\gridline{\fig{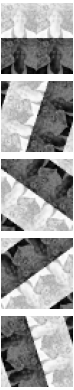}{0.2\textwidth}{}
          \fig{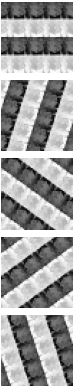}{0.2\textwidth}{}
          \fig{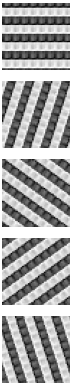}{0.2\textwidth}{}
          \fig{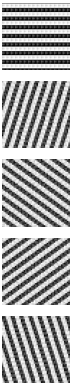}{0.2\textwidth}{}
          \fig{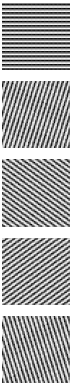}{0.2\textwidth}{}
          }
\caption{
\label{fig:jensenclem}}
\end{figure*}
\newpage

\subsection{Ngo Basis}
\begin{figure*}[b]
\gridline{\fig{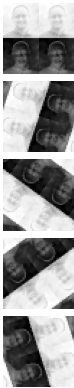}{0.2\textwidth}{}
          \fig{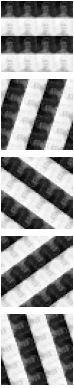}{0.2\textwidth}{}
          \fig{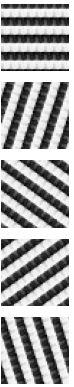}{0.2\textwidth}{}
          \fig{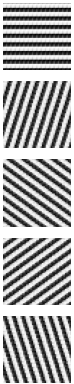}{0.2\textwidth}{}
          \fig{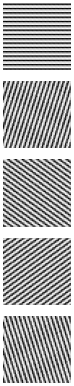}{0.2\textwidth}{}
          }
\caption{
\label{fig:ngo}}
\end{figure*}
\newpage

\subsection{Bowens-Rubin Basis}
\begin{figure*}[b]
\gridline{\fig{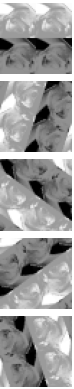}{0.2\textwidth}{}
          \fig{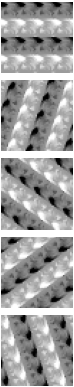}{0.2\textwidth}{}
          \fig{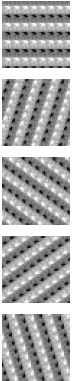}{0.2\textwidth}{}
          \fig{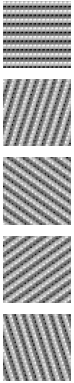}{0.2\textwidth}{}
          \fig{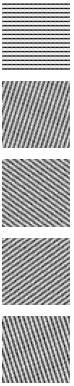}{0.2\textwidth}{}
          }
\caption{
\label{fig:bowensrubin}}
\end{figure*}
\newpage

\subsection{Steiger Basis}
\begin{figure*}[b]
\gridline{\fig{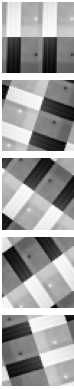}{0.2\textwidth}{}
          \fig{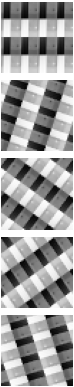}{0.2\textwidth}{}
          \fig{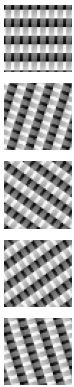}{0.2\textwidth}{}
          \fig{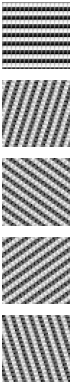}{0.2\textwidth}{}
          \fig{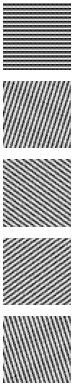}{0.2\textwidth}{}
          }
\caption{
\label{fig:steiger}}
\end{figure*}
\newpage

\subsection{Lewis Basis}
\begin{figure*}[b]
\gridline{\fig{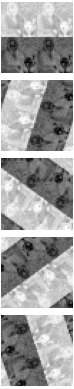}{0.2\textwidth}{}
          \fig{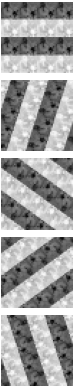}{0.2\textwidth}{}
          \fig{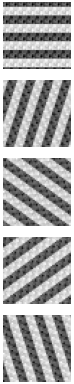}{0.2\textwidth}{}
          \fig{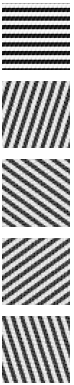}{0.2\textwidth}{}
          \fig{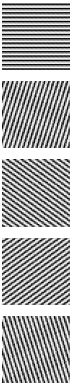}{0.2\textwidth}{}
          }
\caption{
\label{fig:lewis}}
\end{figure*}
\newpage

\subsection{Balmer Basis}
\begin{figure*}[b]
\gridline{\fig{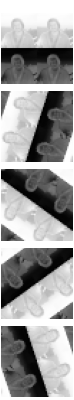}{0.2\textwidth}{}
          \fig{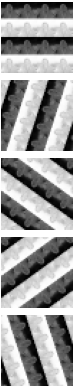}{0.2\textwidth}{}
          \fig{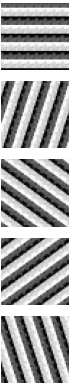}{0.2\textwidth}{}
          \fig{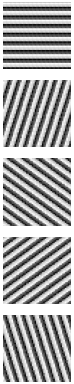}{0.2\textwidth}{}
          \fig{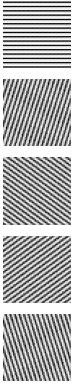}{0.2\textwidth}{}
          }
\caption{
\label{fig:balmer}}
\end{figure*}

\bibliography{hydra}{}
\bibliographystyle{aasjournal}



\end{document}